# An Improved Recursive and Non-recursive Comb Filter for DSP Applications

**Rozita Teymourzadeh** and **Masuri Othman**

Department of Electrical, Electronic and Systems Engineering
VLSI Design Research Group
National University of Malaysia
rozita60@vlsi.eng.ukm.my

**Abstract**

The recursive and non-recursive comb filters are commonly used as decimators for the sigma delta modulators. This paper presents the analysis and design of low power and high speed comb filters. The comparison is made between the recursive and the non-recursive comb filters with the focus on high speed and saving power consumption. Design procedures and examples are given by using Matlab and Verilog HDL for both recursive and non-recursive comb filter with emphasis on frequency response, transfer function and register width. The implementation results show that non-recursive comb filter has capability of speeding up the circuit and reducing power compared to recursive one when the decimation ratio and filter order are high. Using Modified Carry Look-ahead Adder for summation and also apply pipelined filter structure makes it more compatible for DSP application.

## 1 Introduction

Electronic and communication system for speech processing and radar make use of sigma delta modulator in their operation [1], [2]. Future system are required to operate with low power and high speed and therefore the sigma delta modulator must be designed accordingly.

The comb filter that will be designed in this paper is part of the sigma delta modulator which is required to be low power and high speed performance. Comb filter was applied to decimate the sampling frequency from high to low and obtain high resolution. It is also used to perform filtering of the out of band quantization noise and prevent excess aliasing introduced during sampling rate decreasing. Comb filter is popular because of no multipliers and coefficient storages are required due to all filter coefficients are unity [4].

Hence low power and high speed will be key issues in chip implementation of comb decimators.

There are two type of comb filter, first is recursive comb filter and second is non-recursive comb filter.

In 1981, Eugene Hogenauer [3] invented a new class of economical digital filter for decimation called a Cascaded Integrator Comb filter (CIC) or recursive comb filter.

This filter consists of three parts which are Integrator, comb and down sampler. CIC filter is considered as recursive filter because of the feedback loop in integrator circuit.

The second type called non-recursive comb filter which has regular structure and this property makes it suitable for VLSI implementation. In 1999, Yonghong Gao, Lihong Jia and Hannu Tenhunen [9] introduced this filter because of its advantages such as no recursive loop and using low power consumption due to computations are performed at lower sampling rate.

The next section describes the mathematical formulation and block diagram of CIC filters in detail. Section 3 discusses non-recursive comb filter algorithms and its advantages compared to recursive filter. Enhanced high speed architecture is depicted in section 4. Section 5 shows implementation and design result in brief. Finally conclusion is expressed in section 6.

## 2 Recursive CIC filter design

Recursive comb filter or CIC filter consist of N stages of integrator and comb filter which are connected by a down sampler stage as shown in figure 1 in z domain.

The CIC filter has the following transfer function:

$$H(z) = H_I^N(z) \cdot H_C^N(z) = \frac{(1-z^{-RM})^N}{(1-z^{-1})^N} = \left(\sum_{k=0}^{RM-1} z^{-k}\right)^N \quad (1)$$

where N is the number of stage, M is the differential delay and R is the decimation factor.

In this paper, N, M and R have been chosen to be 5, 1 and 16 respectively to avoid overflow in each stages.

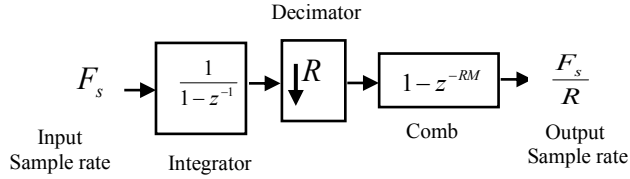

**Figure 1:** *One-stage* of CIC filter block diagram

N, M and R are parameters to determine the register length requirements necessary to assure no data loss. Equation (1) can be express as follow:

$$H(z) = \sum_{k=0}^{(RM-1)N} h(k)z^{-k} = \left[\sum_{k=0}^{RM-1} z^{-k}\right]^N \leq \left|\sum_{k=0}^{RM-1} z^{-k}\right|^N$$

$$\leq \left(\sum_{k=0}^{RM-1} |z|^{-k}\right)^N = \left(\sum_{k=0}^{RM-1} 1\right)^N = (RM)^N \quad (2)$$

From the equation, the maximum register growth/width, $G_{max}$ can be expressed as:

$$G_{max} = (RM)^N \quad (3)$$

In other word, $G_{max}$ is the maximum register growth and a function of the maximum output magnitude due to the worst possible input conditions [3].

If the input data word length is $B_{in}$, most significant bit (MSB) at the filter output, $B_{max}$ is given by:

$$B_{max} = [N\log_2 R + B_{in} - 1] \quad (4)$$

In order to reduce the data loss, normally the first stage of the CIC filter has maximum number of bit compared to the other stages. The recursive part of the CIC filter has to operate with the high oversampling rate and has large width length which is the cause of high power consumption. Due to recursive loop in its structure, low power is limited by the high range of calculation in the integrator stage. Since the integrator stage works at the highest oversampling rate with a large internal word length, decimation ratio and filter order increase which result in more power consumption and speed limitation. In this case, a non-recursive filter is proposed to replace when the decimation ratio and filter order are high.

**2.1 Truncation for speeding up purpose**
Truncation means estimating and removing Least Significant Bit (LSB) to reduce the area requirements on chip and power consumption and also increase speed of calculation. Although this estimation and removing introduces additional error, the error can be made small enough to be acceptable for DSP applications.

Figure 2 illustrates five stages of the CIC filter when $B_{max}$ is 25 bit so truncation is applied to reduce register width. Matlab software helps to find word length in integrator and comb section.

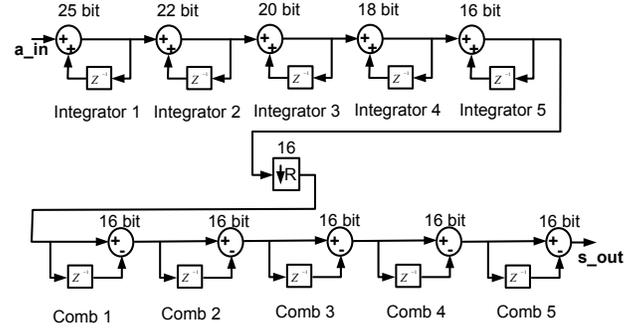

**Figure 2:** Five-stage of truncated CIC filter include integrator and comb cell

## 3 Non-Recursive Comb filter design

The non-recursive comb filter [6], [7] has ability of wide range of rate change. Its transfer function is shown as follow:

$$H(z) = \left(\sum_{i=0}^{R-1} z^{-i}\right)^N = \left(\sum_{i=0}^{2^M-1} z^{-i}\right)^N = \prod_{i=0}^{M-1}\left(1 + z^{-2^i}\right)^N \quad (5)$$

where R is the decimation factor, N is the filter order and M is the number of stage. Note that R should be a power of 2.

Non-recursive comb filter structure is shown in Figure 3. Compared to the CIC filter structure, 1) it is clear that during decimation process and decreasing sampling frequency, the number of bit increases and it is the cause of the saving in power. 2) The comb decimator using the non-recursive algorithm can achieve higher speed since the first stage always has small word length and also 3) when decimation ratio increases, the silicon size of the recursive design algorithm increases slowly compared to the non-recursive design algorithm and it is next advantages of using non-recursive filter as decimator.

Non-recursive comb filter was implemented and R, M and N are selected respectively to be 8, 3 and 5 so if input word length is considered to be 5 output words length change to 20.

Every stage is included of N blocks non-recursive comb filter (See Figure 5).

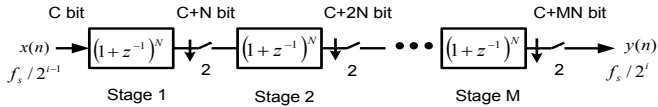

**Figure 3:** Non-recursive Comb decimation filter

## 4 Speed improvement

### 4.1 Pipeline structure

One way to have high speed comb filter is by implementing the pipeline filter structure.

Figure 4 shows pipeline CIC filter structure when truncation is also applied. In the pipelined structure, no additional pipeline registers are used. So that hardware requirement is the same as in the non-pipeline [8].

CIC decimation filter clock rate is determined by the first integrator stage that causes more propagation delay than any other stage due to maximum number of bit. So it is possible to use a higher clock rate for a CIC decimation filter if a pipeline structure is used in the integrator stages, as compared to non-pipelined integrator stages.

Clock rate in integrator section is R times higher than in the comb section, so pipeline structure can not applied for comb section.

Non- recursive comb filter has no integrator part; therefore pipeline structure is used for comb stages.
Figure 5 shows Pipeline non- recursive comb filter. It is achieved by locating register after MCLA in stage M.

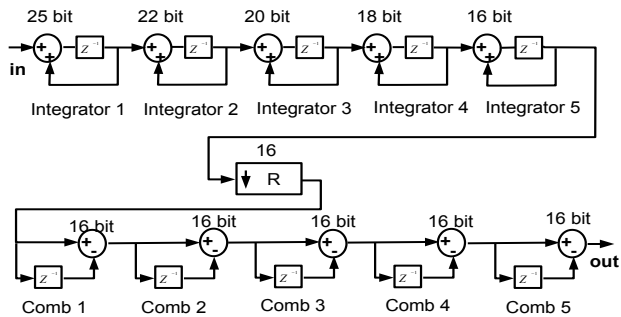

**Figure 4:** Five-stage of truncated pipeline CIC filter include integrator and comb cell

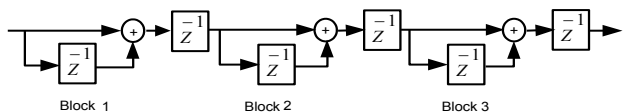

**Figure 5:** Pipelined structure of stage M in non-recursive Comb filter

### 4.2 Modified Carry look-ahead Adder (MCLA)

The other technique to increase speed is using Modified Carry Look-ahead Adder. MCLA was selected to perform the summation in both recursive and non-recursive comb filters.

This improve in speed is due to the carry calculation in MCLA. In the ripple carry adder, most significant bit addition has to wait for the carry to ripple through from the least significant bit addition. Therefore the carry of MCLA adder has become a focus of study in speeding up the adder circuits [5].

The 8 bit MCLA structure is shown in Figure 6. Its block diagram consists of 2, 4-bit module which is connected and each previous 4 bit calculates carry out for the next carry.

The CIC filter in this paper has five MCLA in integrator parts. The maximum number of bit is 25 and it is decreased in next stages. So it truncated respectively to 25, 22, 20, 18 and 16 bit in each adder, left to right.
Non-recursive comb filter has fifteen MCLA in non-recursive part.

The Verilog code has been written to implement summation. The MCLA Verilog code was downloaded to the Xilinx FPGA chip. It was found minimum clock period on FPGA board is 4.389ns (Maximum Frequency is 220 MHz).

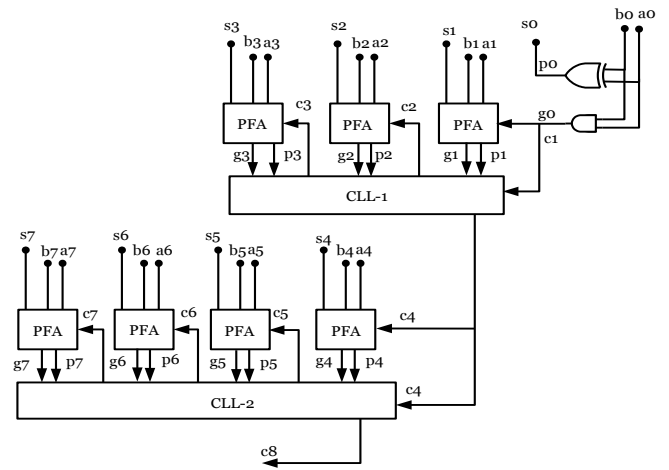

**Figure 6:** The 8 bit MCLA structure

## 5 Design Results

Figure 7 shows the amplitude of the CIC filter output versus output samples number, before and after truncation is applied for the filter.

As seen in Figure 7, output amplitude curve of the filter is sharper when it is truncated and some LSB is discarded compared to filter response without truncation. It shows increasing the speed of filter calculation after truncation.

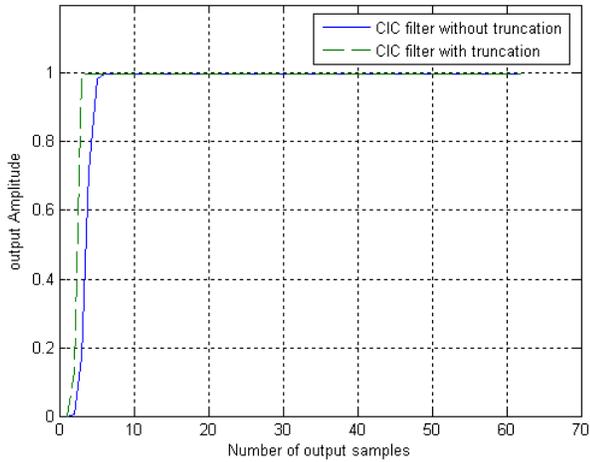

**Figure 7: Truncation effect on CIC filter amplitude vs. output sample number**

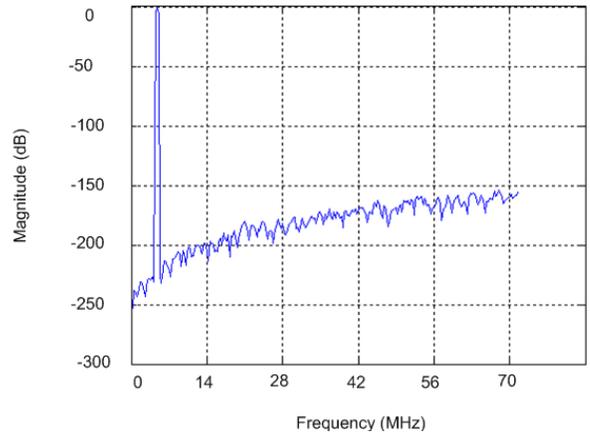

(a)

Figure 8 illustrate the frequency response of the comb filter when the sampling frequency is 6.144 MHz and the pass band frequency is 348 KHz.

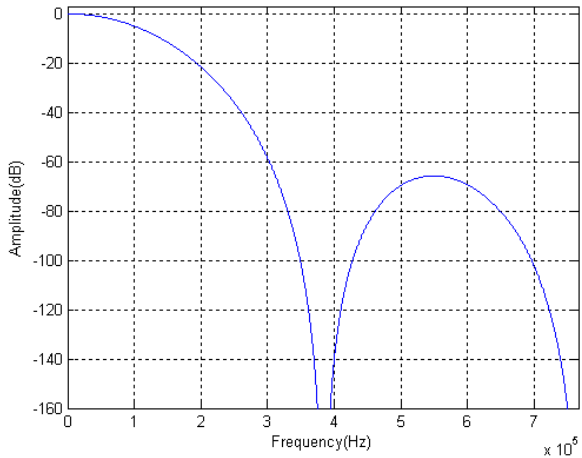

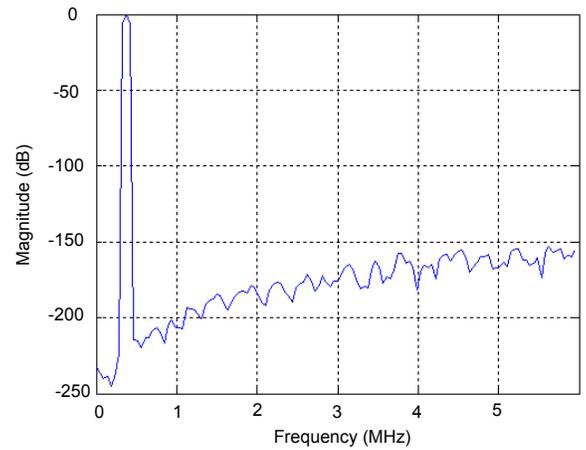

(b)

**Figure 9: Signal spectra (a) Output sigma delta modulator SNR (b) Output Comb filter SNR**

**Figure 8: Comb filter frequency response for *R*=16, *M*=1 and *N*=5**

Figure 9 shows the measured baseband output spectra before (see Figure 9(a)) and after (see Figure 9(b)) the decimation functions. The recursive and non-recursive comb filter Verilog code was wrote and simulated by Matlab software. It is found Signal to Noise ratio (SNR) is 141.56 dB in sigma delta modulator output and SNR is increased to 145.35 dB in the decimation comb stage.

To improve the signal to noise ratio, word length of recursive and non-recursive comb filter should be increased but the speed of filter calculation is also decreased.

Clock frequency versus decimation factor is shown in Figure 10 when highest clock frequency is 90 MHz.

As seen in the figure, recursive comb filter (CIC) curve decreases when decimation factor is increased. According to equation (4), CIC filter word length which has an inverse effect on clock frequency has relation with decimation factor. so increasing decimation factor is the cause of clock frequency limitation. whereas, word length of non recursive comb filter is not depend on decimation factor and increasing decimation factor dose not change clock frequency and its value is constant.

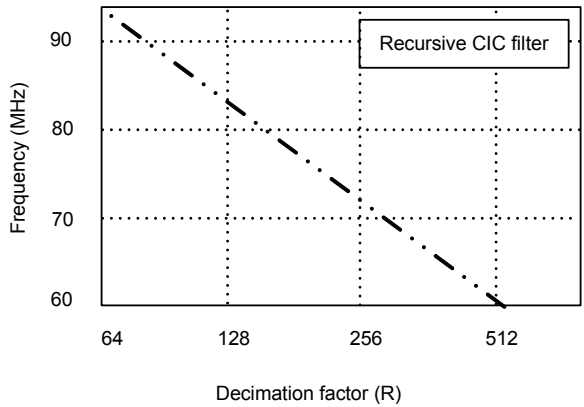

(a)

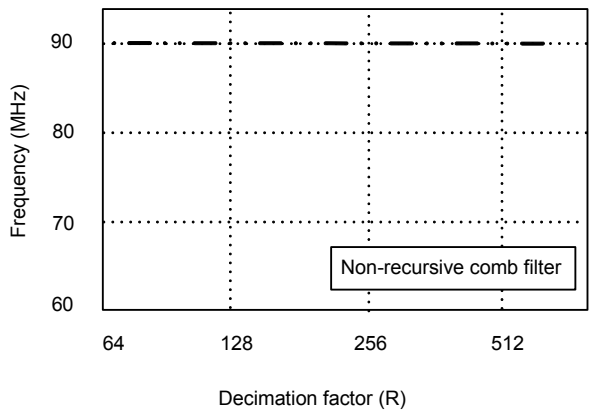

(b)

**Figure 10:** Working frequency comparisons of the (a) recursive and (b) non- recursive comb filter

## 6   Conclusions

Recursive and non-recursive comb filters have been designed and investigated.

Enhanced high Speed recursive and non-recursive comb filters was shown by using pipeline structure and replacing with the modified carry look-ahead adder (MCLA).

The evaluation indicates non-recursive comb filter is attractive compared to the recursive one due to lower power consumption and higher speed.  First stage of recursive comb filter (CIC) word length always has maximum bit number and it is decreased by truncation function whereas first stage of non-recursive filter has minimum word length size compared to other stages, so it is the cause of achieving higher speed.

However Recursive comb filter (CIC) is attractive when the decimation ratio and filter order are not high.